\documentclass[aps,prb, preprint, superscriptaddress, showpacs]{revtex4-1}

\usepackage{graphics}
\usepackage[english]{babel}
\usepackage{epstopdf}
\usepackage{amsmath}
\usepackage{amsfonts}
\usepackage{amssymb}
\usepackage[latin1]{inputenc} 
\usepackage[OT2,T1]{fontenc}

\begin{document}


\title{Control of the gyration dynamics of magnetic vortices by the magnetoelastic effect}



\author{S. Finizio}
\email[Corresponding Author: ]{simone.finizio@psi.ch}
\affiliation{Paul Scherrer Institut, 5232 Villigen PSI, Switzerland}

\author{S. Wintz}
\affiliation{Paul Scherrer Institut, 5232 Villigen PSI, Switzerland}
\affiliation{Helmholtz-Zentrum Dresden-Rossendorf, 01328 Dresden, Germany}

\author{E. Kirk}
\affiliation{Paul Scherrer Institut, 5232 Villigen PSI, Switzerland}
\affiliation{Laboratory for Mesoscopic Systems, Department of Materials, ETH Z\"{u}rich, 8093 Z\"{u}rich, Switzerland}

\author{A. K. Suszka}
\affiliation{Paul Scherrer Institut, 5232 Villigen PSI, Switzerland}
\affiliation{Laboratory for Mesoscopic Systems, Department of Materials, ETH Z\"{u}rich, 8093 Z\"{u}rich, Switzerland}

\author{S. Gliga}
\affiliation{Paul Scherrer Institut, 5232 Villigen PSI, Switzerland}
\affiliation{School of Physics and Astronomy, University of Glasgow, G12 8QQ Glasgow, United Kingdom}

\author{P. Wohlh\"{u}ter}
\affiliation{Paul Scherrer Institut, 5232 Villigen PSI, Switzerland}
\affiliation{Laboratory for Mesoscopic Systems, Department of Materials, ETH Z\"{u}rich, 8093 Z\"{u}rich, Switzerland}

\author{K. Zeissler}
\affiliation{School of Physics and Astronomy, University of Leeds, LS2 9JT Leeds, United Kingdom}

\author{J. Raabe}
\affiliation{Paul Scherrer Institut, 5232 Villigen PSI, Switzerland}

\date{\today}


\begin{abstract}
The influence of a strain-induced uniaxial magnetoelastic anisotropy on the magnetic vortex core dynamics in microstructured magnetostrictive Co$_{40}$Fe$_{40}$B$_{20}$ elements was investigated with time-resolved scanning transmission x-ray microscopy. The measurements revealed a monotonically decreasing eigenfrequency of the vortex core gyration with the increasing magnetoelastic anisotropy, which follows closely the predictions from micromagnetic modeling.
\end{abstract}



\maketitle

\section{Introduction}

Magnetoelectric multiferroics, i.e. materials exhibiting a combination of both ferroelectric and ferromagnetic orders that are coupled with each other, are of prime interest for the development of highly efficient spintronic devices. This is due to the possibility of controlling the magnetization of these systems with an electric field \cite{art:fiebig_multiferroics, art:spaldin_multiferroics}. The extreme rarity of intrinsic magnetoelectric multiferroics \cite{art:spaldin_natural_multiferroics} has however encouraged research on artificial (or composite) multiferroic materials, i.e. composed of different materials that, combined together, exhibit a magnetoelectric coupling \cite{art:fiebig_multiferroics}. An example of such artificial magnetoelectric multiferroic is given by the combination of a piezoelectric and a magnetostrictive material, where the application of an electric field across the piezoelectric leads to the mechanical straining of the magnetostrictive material. Through the magnetoelastic (ME) effect, many different magnetic properties of the magnetostrictive material can be influenced through the application of a mechanical strain \cite{art:lee_magnetostriction}. These effects include e.g. the strain-induced control of the magnetic anisotropy \cite{art:lee_magnetostriction, art:finizio_Ni_PMNPT, art:buzzi_Ni_PMNPT, art:hockel_Ni_PMNPT, art:weiler_PMNPT, art:cui_PZT, art:parkes_galfenol, art:LNO_Ni}, and the control of the saturation magnetization and Curie temperature  \cite{art:liu_vervey_transition, art:adamo_strain_LSMO}.

The ME coupling does not only influence the static magnetic configuration of magnetostrictive materials, but also their magnetodynamical properties. A recent example on the influence of the ME coupling on the magnetodynamical processes of a magnetostrictive material is reported in Ref. \onlinecite{art:LNO_Ni}, where a dynamical strain variation (generated with a surface acoustic wave) was employed for the excitation of microstructured Ni squares at the ns timescale. Also a static mechanical strain is predicted to influence the magnetodynamical response of a magnetostrictive material. According to micromagnetic simulations \cite{art:parkes_galfenol, art:roy_vortexMotionUniax}, the use of the ME coupling could be employed to control the gyration dynamics of magnetic vortices in microstructured magnetostrictive elements through the generation of a ME anisotropy. This is of interest for the fabrication of frequency-tunable spin torque vortex oscillators, as the ME coupling would provide a pathway for an additional control of their eigenfrequency through the application of an electric field \cite{art:parkes_galfenol}. However, up to now, only simulations of the influence of the ME anisotropy on the orbit of the vortex core gyration have been reported. In the work presented here, we provide a first experimental demonstration, using time-resolved x-ray magnetic microscopy, of the influence of the ME coupling on the gyration dynamics of magnetic vortex cores stabilized in magnetostrictive microstructures. We observed a very good agreement between the experimental data and the predictions from micromagnetic modeling, therefore providing an experimental verification that a static strain can be employed to control the magnetodynamical properties of magnetostrictive materials.


\section{Theoretical model and micromagnetic simulations}

The application of an in-plane magnetic field to a microstructured element where a magnetic vortex state is stabilized causes the displacement of the vortex core from its equilibrium position. As the external magnetic field is removed, the vortex relaxes back to its equilibrium position following a damped gyrotropic motion. Such motion is described by Thiele's equation of motion \cite{art:thieleEquation}. Under the assumption of a small displacement of the vortex core from the equilibrium position, and no deformation of the vortex during the excitation, Thiele's equation of motion takes the following form \cite{art:huber_thieleEquation, art:thieleEquation, art:wysin_thieleEquation, art:krueger_harmonicPotential}:
\begin{equation}
 -\mathbf{G} \times \mathbf{\dot{r}} - D \mathbf{\dot{r}} + \mathbf{\nabla} V(\mathbf{r}) \, = \, 0,
 \label{eq:thiele}
\end{equation}
where $\mathbf{G} = G \mathbf{\hat{e}_z}$ is the gyrovector, $\mathbf{r} \, = \, \mathbf{x} + \mathbf{y}$ the displacement of the vortex core with respect to its equilibrium position, and $D$ a damping term related to the Gilbert damping $\alpha$ \cite{art:krueger_harmonicPotential}. The recall force $\mathbf{F}(\mathbf{r}) \, = \, -\mathbf{\nabla} V(\mathbf{r})$ is described with a spring-like term, given in its general form by $\mathbf{F}(\mathbf{r}) \, = \, -\underline{\underline{k}}\mathbf{r}$, where $\underline{\underline{k}}$ is the "spring constant" tensor \cite{art:roy_vortexMotionUniax, art:krueger_harmonicPotential}. Assuming that only the terms $k_{ii}$ are non-vanishing (i.e. the recall force is linear with the displacement of the vortex core \cite{art:roy_vortexMotionUniax}), the recall force can be described as $\mathbf{F}(\mathbf{r}) \, = \, -k_x\mathbf{x} -k_y\mathbf{y}$, defining $k_x \equiv k_{xx}$ and $k_y \equiv k_{yy}$. The potential of the vortex can therefore be written as $V(\mathbf{r}) \, = \, 1/2 (k_x x^2 + k_y y^2)$, which describes a harmonic potential well.

Assuming a negligible Gilbert damping (i.e. $\alpha << 1$, which allows us to neglect the term $- D \mathbf{\dot{r}}$ in Eq. \eqref{eq:thiele}), Thiele's equation of motion leads to an elliptical motion of the vortex core with angular eigenfrequency $\omega_0$ and orbit eccentricity $e$ given by the following relations:
\begin{align}
&\omega_0 \, = \, \frac{\sqrt{k_x k_y}}{G}, \label{eq:eigenfrequency} \\
&e \, = \, \sqrt{\frac{(k_y - k_x)}{k_y}}, \label{eq:eccentricity}
\end{align}
assuming $k_x \leq k_y$ (e.g. when a uniaxial anisotropy is applied along the $x$ direction in an otherwise symmetric structure \cite{art:roy_vortexMotionUniax}). In the simple case of no applied anisotropy ($k_x = k_y = k$), the vortex will exhibit a circular orbit (i.e. $e = 0$) with the expected eigenfrequency of $\omega_0 \, = \, k/G$ \cite{art:wysin_thieleEquation}. If damping is included, then the motion of the vortex can be described by an exponentially damped spiral, where the exponential damping term is related to the Gilbert damping $\alpha$ \cite{art:krueger_harmonicPotential}.

Therefore, it is expected that the straining of a magnetostrictive element in which a magnetic vortex state is stabilized will give rise to a change of the eigenfrequency $\omega_0$ as well as of the eccentricity $e$ of the vortex gyration orbit as a result of the generated uniaxial ME anisotropy K$_\mathrm{ME}$. This prediction is reproduced by micromagnetic simulations of the vortex gyration as a function of an applied uniaxial ME anisotropy.

The motion of the magnetic vortex core in microstructured magnetostrictive Co$_{40}$Fe$_{40}$B$_{20}$ (CoFeB) elements was simulated as a function of an applied ME anisotropy with the MuMax$^3$ finite differences simulation package \cite{art:mumax}. For the micromagnetic simulations, a saturation magnetization of $M_s = 10^6$ A m$^{-1}$, an exchange stiffness of $A = 10^{-11}$ J m$^{-1}$, and a Gilbert damping of $\alpha = 0.01$ were considered. This gives rise to an exchange length of about 4 nm. Therefore, the micromagnetic simulations were carried out with a discretization cell size of 2 nm $\times$ 2 nm. The ME anisotropy was simulated by imposing an external uniaxial anisotropy along the $x$ direction (from 0 to about 12 kJ m$^{-3}$ in steps of 0.5 kJ m$^{-3}$). The magnetic configuration was initialized in the vortex state, and allowed to relax to its equilibrium state before exciting the motion of the vortex core.

The excitation of the motion of the magnetic vortex core was performed by simulating an in-plane pulsed magnetic field along the $y$ direction, to reproduce the experimental conditions. The magnetic field pulses had an amplitude of 2.5 mT and a width of 2 ns. The magnetic configuration of the simulated sample was recorded at intervals of 200 ps, and the simulations were allowed to run for 100 ns after the removal of the magnetic field pulse.

The position of the vortex core was extracted from the micromagnetic simulations by finding the position where the absolute value of the out-of-plane component of the magnetization has its maximum. From this, the eigenfrequency and the eccentricity of the vortex orbit could be extracted. As a non-zero Gilbert damping $\alpha$ is considered in the simulations, the time-dependent position of the vortex core was fitted with an exponentially damped sinusoid. 

\begin{figure}[!hbt]
 \includegraphics{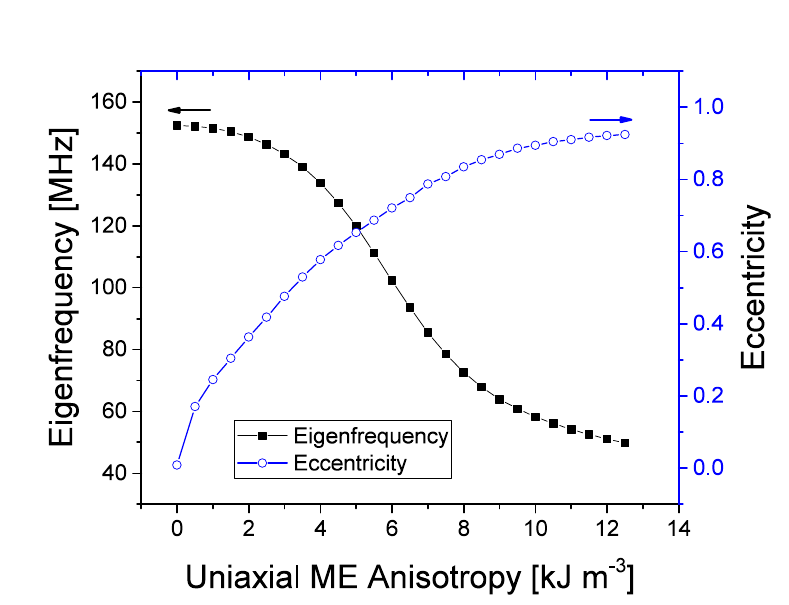}
 \caption{Micromagnetic simulations of the vortex core gyration eigenfrequency (black squares) and the orbit eccentricity (blue circles) for a 35 nm thick, 2 $\mu$m wide CoFeB microstructured square as a function of an applied ME anisotropy.}
 \label{fig:simulations}
\end{figure}

Fig. \ref{fig:simulations} shows the simulated dependence of the eigenfrequency and eccentricity of the vortex orbit as a function of the applied ME anisotropy for the simulated CoFeB microstructures. The simulated eigenfrequency exhibits its maximum value in the case of no applied strain, followed by a monotonous reduction when increasing the anisotropy. On the other hand, the eccentricity is zero for no applied strain (i.e. circular orbit), and increases monotonically with the applied ME anisotropy, underlining the increasingly elliptical vortex core orbit.

\section{Time-resolved imaging of the gyration dynamics}

We verified the predictions both of Thiele's equation of motion and of the micromagnetic simulations by experimentally determining the eigenfrequency of the vortex gyration as a function of the applied ME anisotropy in 35 nm thick, 2 $\mu$m wide, microstructured CoFeB squares. The CoFeB microstructured square elements were fabricated by electron beam lithography on top of 50 nm thick Si$_3$N$_4$ membranes with a rectangular shaped window. Prior to the lithographical exposure, a bilayer of methyl-methacrylate (MMA) and of its copolymer poly-methyl-methacrylate (PMMA) was spun on top of the sample. The lithographical exposure was carried out with a Vistec EBPG 5000Plus electron beam writer, employing electrons at an energy of 100 keV, and an exposure dose of 1800 $\mu$C cm$^{-2}$. A layer of 7 nm of Al was deposited by thermal evaporation on top of the resist to reduce charging effects in the electron beam writer. After the exposure, the Al layer was removed by immersion in tetramethylammonium-hydroxide followed by rinsing in deionized water, and the resist was developed by immersion in a solution of methyl-isobutyl-ketone 1:3 in isopropanol (IPA) for 120 s followed by immersion in pure IPA for 60 s.

The CoFeB films were deposited by DC sputtering (with a growth rate of 1.7 nm min$^{-1}$) in a dedicated chamber with a base pressure in the $10^{-9}$ mbar range at an Ar pressure of about 4 mbar. The films were then capped with 4 nm of Ta. After deposition, the unexposed resist was removed together with the metal film on top by immersion in pure acetone (lift-off).

To excite the motion of the magnetic vortices, a Cu stripline was fabricated on top of the CoFeB microstructures, allowing for the generation of an in-plane magnetic field through the Oersted effect. For the fabrication of the Cu stripline, a bilayer of MMA and PMMA was spun on top of the sample, and a lithographical exposure with an electron energy of 100 keV and a dose of 1500 $\mu$C cm$^{-2}$ was carried out. To reduce charging, a 7 nm thick Al layer was once again deposited on top of the resist prior to the lithographical exposure. The samples were then developed using the same recipe employed for the CoFeB step. A 100 nm thick Cu layer was deposited by thermal evaporation in a dedicated chamber with a base pressure in the $10^{-7}$ mbar range.

\begin{figure}
 \includegraphics{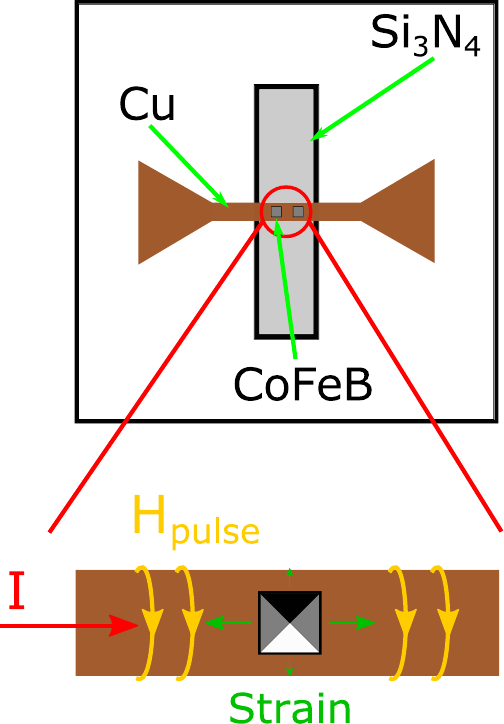}
 \caption{Sketch of the samples employed for the TR-STXM investigations. The CoFeB microstructured elements were fabricated on a rectangular Si$_3$N$_4$ membrane, which was employed for the mechanical straining of the microstructured elements. A Cu stripline was fabricated on top of the CoFeB elements to excite, through the generation of an in-plane Oersted field, the gyration of the magnetic vortices in the CoFeB elements.}
 \label{fig:sample}
\end{figure}

A sketch of the samples employed for the investigations presented here, showing both the CoFeB microstructured elements and the Cu stripline, is shown in Fig. \ref{fig:sample}.

The magnetic configuration of the CoFeB microstructured elements was investigated with time-resolved scanning transmission x-ray microscopy (TR-STXM) imaging. Circularly-polarized x-rays tuned to the Co L$_3$ absorption edge were employed to detect the magnetic configuration of the CoFeB squares exploiting the x-ray magnetic circular dichroism (XMCD) effect. The samples were mounted with their surface oriented at 30$^\circ$ with respect to the incoming x-ray beam, therefore allowing for the imaging of the in-plane component of the magnetization. To compensate for the 30$^\circ$ orientation of the sample with respect to the x-ray beam, the images were scaled with a factor $2/\sqrt{3}$ in the $x$ direction. For the data presented here, the spatial resolution was 25 nm.

\begin{figure*}
 \includegraphics{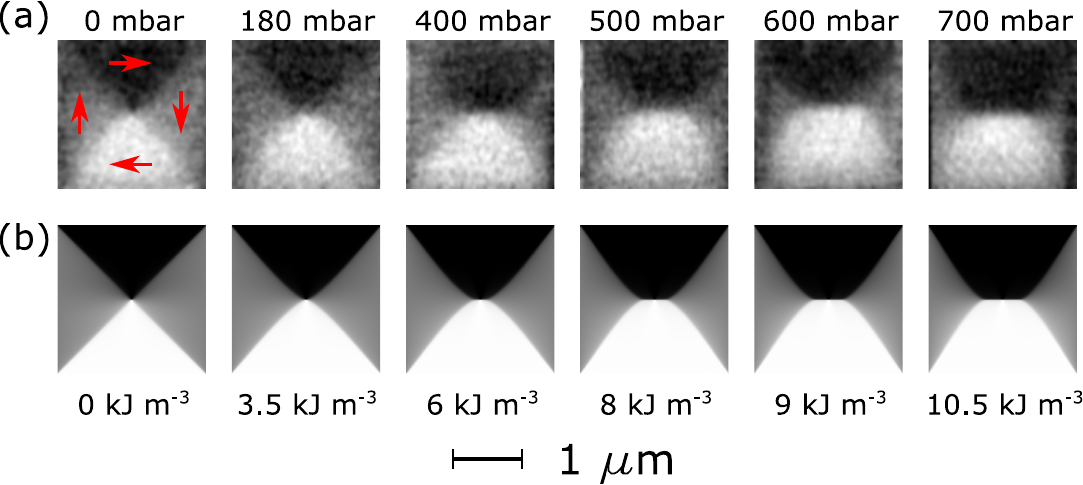}
 \caption{(a) XMCD-STXM images of a 2 $\mu$m wide CoFeB square fabricated on a thin Si$_3$N$_4$ membrane as a function of the pressure difference between the two sides of the membrane, showing the generation of a uniaxial ME anisotropy caused by the bending of the membrane. (b) Micromagnetic simulations of the CoFeB square at different applied ME anisotropies. By comparing the XMCD-STXM images with the corresponding micromagnetic simulations, the magnitude of the ME anisotropy can be extracted. The red arrows indicate the direction of the magnetization in the images.}
 \label{fig:quasi-static}
\end{figure*}

The CoFeB microstructured elements were strained by mechanically bending the Si$_3$N$_4$ membrane with the application of a pressure difference between the two sides of the membrane \cite{art:bending_setup}. The application of a non-isotropic strain to the magnetostrictive CoFeB microstructures (stemming from the use of rectangular shaped membranes \cite{art:bending_setup}) leads to the generation of a uniaxial ME anisotropy, as shown in Fig. \ref{fig:quasi-static}. The measurements of the static response of the magnetization as a function of the applied strain shown in Fig. \ref{fig:quasi-static} are in agreement with the results from previous works reported for different magnetostrictive materials \cite{art:finizio_Ni_PMNPT, art:buzzi_Ni_PMNPT, art:weiler_PMNPT, art:hockel_Ni_PMNPT}. 


The gyration of the magnetic vortices was investigated by TR-STXM imaging, using photons of only one helicity (circular negative). Here, the time resolved images were acquired in a pump-probe scheme. The pump signal consisted of a pulsed in-plane magnetic field, generated by the injection of a current across the Cu stripline, generated with a Tektronix AWG7122C arbitrary waveform generator. The probing signal is given by the x-ray flashes generated by the synchrotron light source. For the data presented here, the temporal resolution was 400 ps.

The waveform generator was synchronized to the master clock of the synchrotron, allowing for the synchronization between the pump and probing signals. The synchronization signal is generated by a dedicated field programmable gate array from the master clock of the synchrotron, which also handles the acquisition of the time-resolved data. Thereby it is guaranteed that, for each of the time-resolved scans (whilst maintaining the same number of frames and the same temporal resolution), the zero position of the time delay occurs at the same temporal position. 

The width of the current pulses injected across the Cu stripline was selected to be between 2 and 6 ns, with two consecutive pulses separated by about 150-200 ns, to allow both for the reduction of Joule heating effects caused by the injection of the current pulses and to allow for the dynamical processes to end before the next excitation. The shape of the current pulses, which was employed as a diagnostic for the status of the stripline, was monitored before (through a -20 dB pick-off T) and after the sample with an Agilent Infiniium DSO-S 404A real time oscilloscope.

\begin{figure*}
 \includegraphics{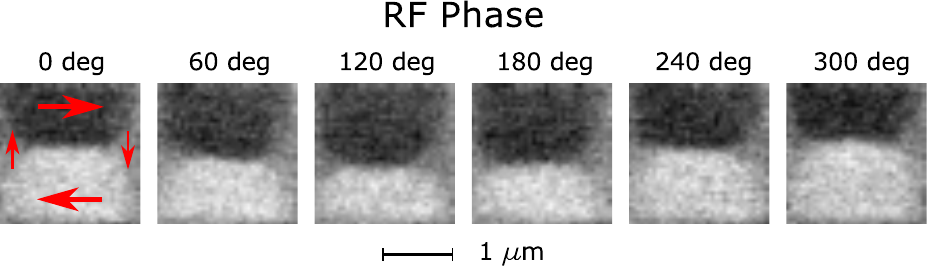}
 \caption{Snapshots of a TR-STXM image of the gyration dynamics of the magnetic vortex in a 25 nm thick, 2 $\mu$m wide CoFeB microstructured square at different phases of the excitation signal. The vortex was excited by generating a 47 MHz sinusoidal magnetic field. The red arrows indicate the direction of the magnetization in the images.}
 \label{fig:tr_xmcd}
\end{figure*}

An example of a TR-STXM image of the vortex gyration dynamics in a 25 nm thick CoFeB microstructured element is shown in Fig. \ref{fig:tr_xmcd}. Here, a sinusoidal magnetic field with its frequency tuned close to the gyration eigenfrequency was employed.

\begin{figure*}
 \includegraphics{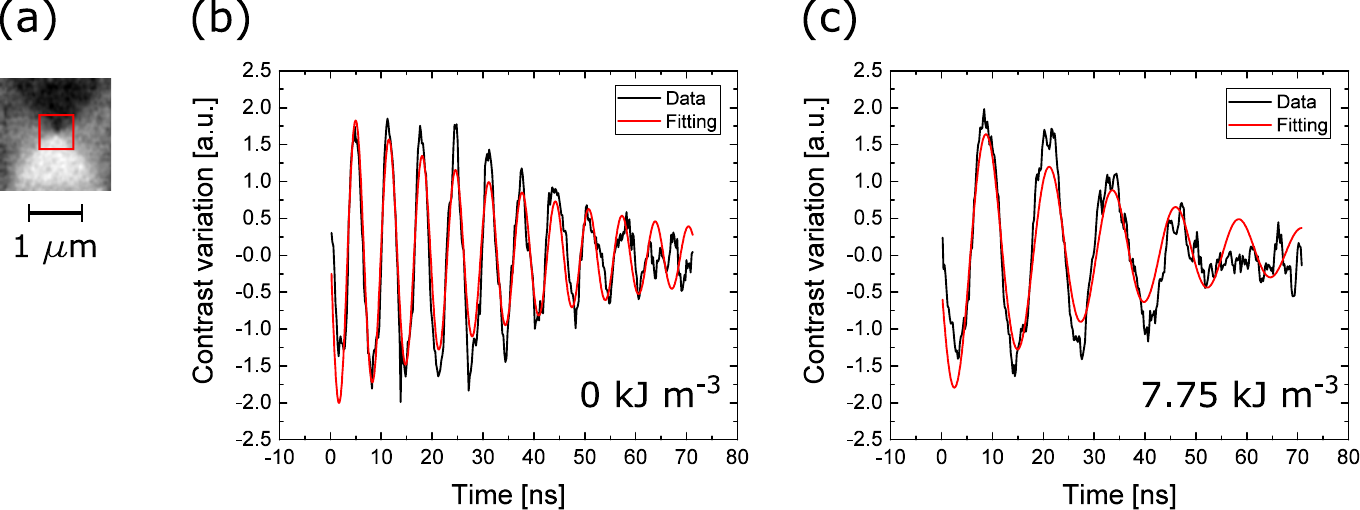}
 \caption{Fitting of the magnetic contrast variation determined from the time-resolved STXM images. (a) Region of interest (750 $\times$ 750 nm$^2$), centered on the equilibrium position of the magnetic vortex core. (b-c) Fit of the time-resolved XMCD contrast variation in the region of interest shown in (a) at an applied uniaxial anisotropy of (b) 0 kJ m$^{-3}$ and (c) of about 7.75 kJ m$^{-3}$. The experimental data points were fitted with an exponentially damped sinusoid.}
 \label{fig:fitting_experiments}
\end{figure*}

The eigenfrequency was determined by measuring the time-resolved variation in the XMCD contrast in a region of interest centered around the equilibrium position of the magnetic vortex (see Fig. \ref{fig:fitting_experiments}(a)) upon the application of a magnetic field pulse, and by fitting it with an exponentially damped sinusoid (see Fig. \ref{fig:fitting_experiments}(b-c) for an example of such fitting). Prior to the acquisition of the time-resolved images, the magnitude of the applied ME anisotropy was measured by acquiring static XMCD-STXM images of the magnetic configuration of the microstructured square at different applied strains (i.e. pressure differences between the two sides of the Si$_3$N$_4$ membrane). The images were compared with micromagnetic simulations of the magnetic configuration at different ME anisotropies (see Fig. \ref{fig:quasi-static}), which provides a reliable method for determining the magnitude of the ME anisotropy \cite{art:finizio_Ni_PMNPT}.

\begin{figure}[!hbt]
 \includegraphics{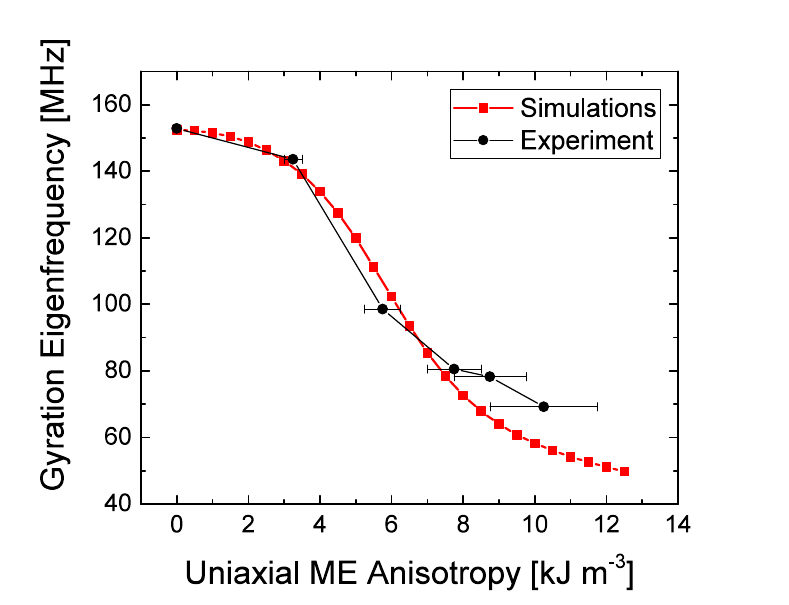}
 \caption{Comparison between the experimental measurements of the gyration eigenfrequency in a 2 $\mu$m wide, 35 nm thick, CoFeB square as a function of the applied ME anisotropy and the eigenfrequency predicted from micromagnetic simulations.}
 \label{fig:pulsed}
\end{figure}

The experimental measurements of the gyration eigenfrequency as a function of the ME anisotropy, compared with the results from the micromagnetic simulations, are shown in Fig. \ref{fig:pulsed}. Here, we observed that the experimental data closely follows the predicted behavior from the micromagnetic simulations. A maximum in the eigenfrequency for no applied strain (i.e. no applied ME anisotropy), followed by a monotonic reduction of the eigenfrequency with the increase of the applied ME anisotropy was observed.

These results provide a clear demonstration that the straining of magnetostrictive microstructured elements can be employed to modify the potential well of a magnetic vortex, and therefore to affect the eigenfrequency of the vortex gyration through the ME coupling. Furthermore, it was verified that the process is reproducible and reversible by removing and re-applying the strain to the CoFeB squares.

As shown in Fig. \ref{fig:pulsed}, it is possible to obtain a decrease of almost a factor of 3 in the vortex core gyration eigenfrequency with applied ME anisotropies on the order of 10 kJ m$^{-3}$. These values of the ME anisotropy are easily achievable (with applied strains on the order of 100 to 1000 ppm \cite{art:bending_setup}) in magnetostrictive materials such as CoFeB, Ni, and Ga$_x$Fe$_{1-x}$, i.e. materials with magnetostrictive constants on the order of 10 to 100 ppm \cite{art:finizio_Ni_PMNPT, art:buzzi_Ni_PMNPT, art:weiler_PMNPT, art:parkes_galfenol}.

\begin{figure}[!hbt]
 \includegraphics{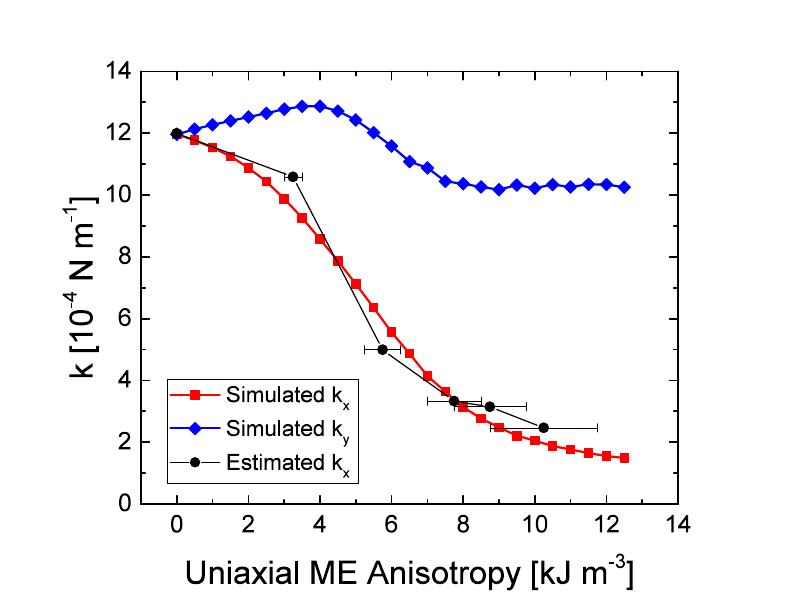}
 \caption{Values of the "spring constants" $k_x$ and $k_y$ determined from micromagnetic simulations, and estimation of the $k_x$ value from the experimentally-determined gyration eigenfrequency under the assumption of a constant $k_y$.}
 \label{fig:potential}
\end{figure}

The very good agreement between the experimental data and the micromagnetic simulations allows us to reliably describe the potential well $V(\mathbf{r}) \, = \, 1/2 (k_x x^2 + k_y y^2)$ of the magnetic vortex core. In particular, from the simulated values of the gyration eigenfrequency and orbit eccentricity, it is possible to extract the two "spring constants" $k_x$ and k$_y$. Their predicted dependence on the ME anisotropy is shown in Fig. \ref{fig:potential}.

In particular, we observe that, for the considered values of the ME anisotropy, the harmonic potential of the vortex loses stiffness along the direction of the applied anisotropy (i.e. $k_x$ is monotonically decreasing with the applied anisotropy), whilst retaining about the same stiffness along the direction perpendicular to the applied anisotropy (i.e. $k_y$ is almost constant in the range of anisotropies considered here). This given, the value of $k_x$ as a function of the applied anisotropy can be estimated from the measured values of the gyration eigenfrequency with the following relation, calculated from Eq. \eqref{eq:eigenfrequency} and \eqref{eq:eccentricity} with the assumption of a constant $k_y$:
\begin{equation}
k_x(K_\mathrm{ME}) \, \sim \, G\frac{\omega_0^2(K_\mathrm{ME})}{\omega_0(K_\mathrm{ME}=0)}. 
\label{eq:estimation}
\end{equation}
The values of $k_x$ estimated using Eq. \eqref{eq:estimation} from the experimental data are shown in Fig. \ref{fig:potential}, demonstrating that this estimation yields a reasonable approximation of the stiffness of the vortex potential well for low values of the ME anisotropy, where $k_y$ can be assumed as constant. This allows the estimation of the stiffness of the potential well with the measurement of only the eigenfrequency of the vortex gyration, which can be easily determined from the time-resolved STXM images.

Note that, for higher values of the applied anisotropy, also $k_y$ will eventually start to decrease with respect to the $K_\mathrm{ME} \, = \, 0$ case, following then the trend simulated for Ga$_x$Fe$_{1-x}$ microstructured elements in Ref. \onlinecite{art:roy_vortexMotionUniax}.


\section{Conclusions}

In conclusion, we reported on the manipulation of vortex core dynamics in magnetostrictive CoFeB microstructures through the ME coupling. We have observed that, when generating a uniaxial ME anisotropy in the CoFeB microstructures through the ME coupling, the harmonic potential of the vortex core responds to the applied anisotropy via a decrease in the stiffness of the potential along the direction of the applied anisotropy. This affects the dynamics of the magnetic vortex core in two ways: the gyration eigenfrequency of the vortex core decreases with increasing anisotropy, and the orbit of the vortex core becomes elliptical, with the major axis of the orbit being oriented along the applied anisotropy axis. The measured changes of the vortex core eigenfrequency follow the predictions from the micromagnetic model closely. Furthermore, it was possible to employ the measured vortex core eigenfrequency to estimate the changes in the stiffness of the potential well of the vortex core induced by the ME coupling. We therefore demonstrated experimentally that the ME coupling can be employed for the reliable and reproducible control of the magnetodynamical processes in magnetostrictive materials.


\begin{acknowledgments}
This work was performed at the PolLux (X07DA) endstation of the Swiss Light Source, Paul Scherrer Institut, Villigen, Switzerland. The authors would like to thank B. Sarafimov for technical support. The research leading to these results has received funding from the European Community's Seventh Framework Programme (FP7/2007-2013) under grant agreement No. 290605 (PSI-FELLOW/COFUND), and the European Union's Horizon 2020 Project MAGicSky (Grant No. 665095). The PolLux endstation was financed by the German Minister f\"ur Bildung und Forschung (BMBF) through contracts 05KS4WE1/6 and 05KS7WE1. SG was funded by the European Union's Horizon 2020 Research and Innovation Programme under the Marie Sklodowska-Curie grant agreement No. 708674.
\end{acknowledgments}


\begin{thebibliography}{21}%
\makeatletter
\providecommand \@ifxundefined [1]{%
 \@ifx{#1\undefined}
}%
\providecommand \@ifnum [1]{%
 \ifnum #1\expandafter \@firstoftwo
 \else \expandafter \@secondoftwo
 \fi
}%
\providecommand \@ifx [1]{%
 \ifx #1\expandafter \@firstoftwo
 \else \expandafter \@secondoftwo
 \fi
}%
\providecommand \natexlab [1]{#1}%
\providecommand \enquote  [1]{``#1''}%
\providecommand \bibnamefont  [1]{#1}%
\providecommand \bibfnamefont [1]{#1}%
\providecommand \citenamefont [1]{#1}%
\providecommand \href@noop [0]{\@secondoftwo}%
\providecommand \href [0]{\begingroup \@sanitize@url \@href}%
\providecommand \@href[1]{\@@startlink{#1}\@@href}%
\providecommand \@@href[1]{\endgroup#1\@@endlink}%
\providecommand \@sanitize@url [0]{\catcode `\\12\catcode `\$12\catcode
  `\&12\catcode `\#12\catcode `\^12\catcode `\_12\catcode `\%12\relax}%
\providecommand \@@startlink[1]{}%
\providecommand \@@endlink[0]{}%
\providecommand \url  [0]{\begingroup\@sanitize@url \@url }%
\providecommand \@url [1]{\endgroup\@href {#1}{\urlprefix }}%
\providecommand \urlprefix  [0]{URL }%
\providecommand \Eprint [0]{\href }%
\providecommand \doibase [0]{http://dx.doi.org/}%
\providecommand \selectlanguage [0]{\@gobble}%
\providecommand \bibinfo  [0]{\@secondoftwo}%
\providecommand \bibfield  [0]{\@secondoftwo}%
\providecommand \translation [1]{[#1]}%
\providecommand \BibitemOpen [0]{}%
\providecommand \bibitemStop [0]{}%
\providecommand \bibitemNoStop [0]{.\EOS\space}%
\providecommand \EOS [0]{\spacefactor3000\relax}%
\providecommand \BibitemShut  [1]{\csname bibitem#1\endcsname}%
\let\auto@bib@innerbib\@empty
\bibitem [{\citenamefont {Fiebig}(2005)}]{art:fiebig_multiferroics}%
  \BibitemOpen
  \bibfield  {author} {\bibinfo {author} {\bibfnamefont {M.}~\bibnamefont
  {Fiebig}},\ }\href@noop {} {\bibfield  {journal} {\bibinfo  {journal}
  {Journal of Physics D: Applied Physics}\ }\textbf {\bibinfo {volume} {38}},\
  \bibinfo {pages} {R123} (\bibinfo {year} {2005})}\BibitemShut {NoStop}%
\bibitem [{\citenamefont {Spaldin}\ and\ \citenamefont
  {Fiebig}(2005)}]{art:spaldin_multiferroics}%
  \BibitemOpen
  \bibfield  {author} {\bibinfo {author} {\bibfnamefont {N.}~\bibnamefont
  {Spaldin}}\ and\ \bibinfo {author} {\bibfnamefont {M.}~\bibnamefont
  {Fiebig}},\ }\href@noop {} {\bibfield  {journal} {\bibinfo  {journal}
  {Science}\ }\textbf {\bibinfo {volume} {309}},\ \bibinfo {pages} {391}
  (\bibinfo {year} {2005})}\BibitemShut {NoStop}%
\bibitem [{\citenamefont {Hill}(2000)}]{art:spaldin_natural_multiferroics}%
  \BibitemOpen
  \bibfield  {author} {\bibinfo {author} {\bibfnamefont {N.}~\bibnamefont
  {Hill}},\ }\href@noop {} {\bibfield  {journal} {\bibinfo  {journal} {Journal
  of Physical Chemistry B}\ }\textbf {\bibinfo {volume} {104}},\ \bibinfo
  {pages} {6694} (\bibinfo {year} {2000})}\BibitemShut {NoStop}%
\bibitem [{\citenamefont {Lee}(1955)}]{art:lee_magnetostriction}%
  \BibitemOpen
  \bibfield  {author} {\bibinfo {author} {\bibfnamefont {E.~W.}\ \bibnamefont
  {Lee}},\ }\href@noop {} {\bibfield  {journal} {\bibinfo  {journal} {Reports
  on Progress in Physics}\ }\textbf {\bibinfo {volume} {18}},\ \bibinfo {pages}
  {184} (\bibinfo {year} {1955})}\BibitemShut {NoStop}%
\bibitem [{\citenamefont {Finizio}\ \emph {et~al.}(2014)\citenamefont
  {Finizio}, \citenamefont {Foerster}, \citenamefont {Buzzi}, \citenamefont
  {Kr{\"u}ger}, \citenamefont {Jourdan}, \citenamefont {Vaz}, \citenamefont
  {Hockel}, \citenamefont {Miyawaki}, \citenamefont {Tkach}, \citenamefont
  {Valencia}, \citenamefont {Kronast}, \citenamefont {Carman}, \citenamefont
  {Nolting},\ and\ \citenamefont {Kl{\"a}ui}}]{art:finizio_Ni_PMNPT}%
  \BibitemOpen
  \bibfield  {author} {\bibinfo {author} {\bibfnamefont {S.}~\bibnamefont
  {Finizio}}, \bibinfo {author} {\bibfnamefont {M.}~\bibnamefont {Foerster}},
  \bibinfo {author} {\bibfnamefont {M.}~\bibnamefont {Buzzi}}, \bibinfo
  {author} {\bibfnamefont {B.}~\bibnamefont {Kr{\"u}ger}}, \bibinfo {author}
  {\bibfnamefont {M.}~\bibnamefont {Jourdan}}, \bibinfo {author} {\bibfnamefont
  {C.}~\bibnamefont {Vaz}}, \bibinfo {author} {\bibfnamefont {J.}~\bibnamefont
  {Hockel}}, \bibinfo {author} {\bibfnamefont {T.}~\bibnamefont {Miyawaki}},
  \bibinfo {author} {\bibfnamefont {A.}~\bibnamefont {Tkach}}, \bibinfo
  {author} {\bibfnamefont {S.}~\bibnamefont {Valencia}}, \bibinfo {author}
  {\bibfnamefont {F.}~\bibnamefont {Kronast}}, \bibinfo {author} {\bibfnamefont
  {G.}~\bibnamefont {Carman}}, \bibinfo {author} {\bibfnamefont
  {F.}~\bibnamefont {Nolting}}, \ and\ \bibinfo {author} {\bibfnamefont
  {M.}~\bibnamefont {Kl{\"a}ui}},\ }\href@noop {} {\bibfield  {journal}
  {\bibinfo  {journal} {Physical Review Applied}\ }\textbf {\bibinfo {volume}
  {1}},\ \bibinfo {pages} {021001} (\bibinfo {year} {2014})}\BibitemShut
  {NoStop}%
\bibitem [{\citenamefont {Buzzi}\ \emph {et~al.}(2013)\citenamefont {Buzzi},
  \citenamefont {Chopdekar}, \citenamefont {Hockel}, \citenamefont {Bur},
  \citenamefont {Wu}, \citenamefont {Pilet}, \citenamefont {Warnicke},
  \citenamefont {Carman}, \citenamefont {Heyderman},\ and\ \citenamefont
  {Nolting}}]{art:buzzi_Ni_PMNPT}%
  \BibitemOpen
  \bibfield  {author} {\bibinfo {author} {\bibfnamefont {M.}~\bibnamefont
  {Buzzi}}, \bibinfo {author} {\bibfnamefont {R.~V.}\ \bibnamefont
  {Chopdekar}}, \bibinfo {author} {\bibfnamefont {J.~L.}\ \bibnamefont
  {Hockel}}, \bibinfo {author} {\bibfnamefont {A.}~\bibnamefont {Bur}},
  \bibinfo {author} {\bibfnamefont {T.}~\bibnamefont {Wu}}, \bibinfo {author}
  {\bibfnamefont {N.}~\bibnamefont {Pilet}}, \bibinfo {author} {\bibfnamefont
  {P.}~\bibnamefont {Warnicke}}, \bibinfo {author} {\bibfnamefont {G.~P.}\
  \bibnamefont {Carman}}, \bibinfo {author} {\bibfnamefont {L.~J.}\
  \bibnamefont {Heyderman}}, \ and\ \bibinfo {author} {\bibfnamefont
  {F.}~\bibnamefont {Nolting}},\ }\href@noop {} {\bibfield  {journal} {\bibinfo
   {journal} {Physical Review Letters}\ }\textbf {\bibinfo {volume} {111}},\
  \bibinfo {pages} {027204} (\bibinfo {year} {2013})}\BibitemShut {NoStop}%
\bibitem [{\citenamefont {Hockel}\ \emph {et~al.}(2012)\citenamefont {Hockel},
  \citenamefont {Bur}, \citenamefont {Wu}, \citenamefont {Wetzlar},\ and\
  \citenamefont {Carman}}]{art:hockel_Ni_PMNPT}%
  \BibitemOpen
  \bibfield  {author} {\bibinfo {author} {\bibfnamefont {J.~L.}\ \bibnamefont
  {Hockel}}, \bibinfo {author} {\bibfnamefont {A.}~\bibnamefont {Bur}},
  \bibinfo {author} {\bibfnamefont {T.}~\bibnamefont {Wu}}, \bibinfo {author}
  {\bibfnamefont {K.~P.}\ \bibnamefont {Wetzlar}}, \ and\ \bibinfo {author}
  {\bibfnamefont {G.~P.}\ \bibnamefont {Carman}},\ }\href@noop {} {\bibfield
  {journal} {\bibinfo  {journal} {Applied Physics Letters}\ }\textbf {\bibinfo
  {volume} {100}},\ \bibinfo {pages} {022401} (\bibinfo {year}
  {2012})}\BibitemShut {NoStop}%
\bibitem [{\citenamefont {Weiler}\ \emph {et~al.}(2009)\citenamefont {Weiler},
  \citenamefont {Brandlmaier}, \citenamefont {Gepr{\"a}gs}, \citenamefont
  {Althammer}, \citenamefont {Opel}, \citenamefont {Bihler}, \citenamefont
  {Huebl}, \citenamefont {Brandt}, \citenamefont {Gross},\ and\ \citenamefont
  {Goennenwein}}]{art:weiler_PMNPT}%
  \BibitemOpen
  \bibfield  {author} {\bibinfo {author} {\bibfnamefont {M.}~\bibnamefont
  {Weiler}}, \bibinfo {author} {\bibfnamefont {A.}~\bibnamefont {Brandlmaier}},
  \bibinfo {author} {\bibfnamefont {S.}~\bibnamefont {Gepr{\"a}gs}}, \bibinfo
  {author} {\bibfnamefont {M.}~\bibnamefont {Althammer}}, \bibinfo {author}
  {\bibfnamefont {M.}~\bibnamefont {Opel}}, \bibinfo {author} {\bibfnamefont
  {C.}~\bibnamefont {Bihler}}, \bibinfo {author} {\bibfnamefont
  {H.}~\bibnamefont {Huebl}}, \bibinfo {author} {\bibfnamefont {M.~S.}\
  \bibnamefont {Brandt}}, \bibinfo {author} {\bibfnamefont {R.}~\bibnamefont
  {Gross}}, \ and\ \bibinfo {author} {\bibfnamefont {S.~T.~B.}\ \bibnamefont
  {Goennenwein}},\ }\href@noop {} {\bibfield  {journal} {\bibinfo  {journal}
  {New Journal of Physics}\ }\textbf {\bibinfo {volume} {11}},\ \bibinfo
  {pages} {013021} (\bibinfo {year} {2009})}\BibitemShut {NoStop}%
\bibitem [{\citenamefont {Cui}\ \emph {et~al.}(2015)\citenamefont {Cui},
  \citenamefont {Liang}, \citenamefont {Paisley}, \citenamefont {Sepulveda},
  \citenamefont {Ihlefeld}, \citenamefont {Carman},\ and\ \citenamefont
  {Lynch}}]{art:cui_PZT}%
  \BibitemOpen
  \bibfield  {author} {\bibinfo {author} {\bibfnamefont {J.}~\bibnamefont
  {Cui}}, \bibinfo {author} {\bibfnamefont {C.~Y.}\ \bibnamefont {Liang}},
  \bibinfo {author} {\bibfnamefont {A.}~\bibnamefont {Paisley}}, \bibinfo
  {author} {\bibfnamefont {A.}~\bibnamefont {Sepulveda}}, \bibinfo {author}
  {\bibfnamefont {J.~F.}\ \bibnamefont {Ihlefeld}}, \bibinfo {author}
  {\bibfnamefont {G.~P.}\ \bibnamefont {Carman}}, \ and\ \bibinfo {author}
  {\bibfnamefont {C.~S.}\ \bibnamefont {Lynch}},\ }\href@noop {} {\bibfield
  {journal} {\bibinfo  {journal} {Applied Physics Letters}\ }\textbf {\bibinfo
  {volume} {107}},\ \bibinfo {pages} {092903} (\bibinfo {year}
  {2015})}\BibitemShut {NoStop}%
\bibitem [{\citenamefont {Parkes}\ \emph {et~al.}(2014)\citenamefont {Parkes},
  \citenamefont {Beardsley}, \citenamefont {Bowe}, \citenamefont {Isakov},
  \citenamefont {Warburton}, \citenamefont {Edmonds}, \citenamefont {Campion},
  \citenamefont {Gallagher}, \citenamefont {Rushforth},\ and\ \citenamefont
  {Cavill}}]{art:parkes_galfenol}%
  \BibitemOpen
  \bibfield  {author} {\bibinfo {author} {\bibfnamefont {D.~E.}\ \bibnamefont
  {Parkes}}, \bibinfo {author} {\bibfnamefont {R.}~\bibnamefont {Beardsley}},
  \bibinfo {author} {\bibfnamefont {S.}~\bibnamefont {Bowe}}, \bibinfo {author}
  {\bibfnamefont {I.}~\bibnamefont {Isakov}}, \bibinfo {author} {\bibfnamefont
  {P.~A.}\ \bibnamefont {Warburton}}, \bibinfo {author} {\bibfnamefont {K.~W.}\
  \bibnamefont {Edmonds}}, \bibinfo {author} {\bibfnamefont {R.~P.}\
  \bibnamefont {Campion}}, \bibinfo {author} {\bibfnamefont {B.~L.}\
  \bibnamefont {Gallagher}}, \bibinfo {author} {\bibfnamefont {A.~W.}\
  \bibnamefont {Rushforth}}, \ and\ \bibinfo {author} {\bibfnamefont {S.~A.}\
  \bibnamefont {Cavill}},\ }\href@noop {} {\bibfield  {journal} {\bibinfo
  {journal} {Applied Physics Letters}\ }\textbf {\bibinfo {volume} {105}},\
  \bibinfo {pages} {062405} (\bibinfo {year} {2014})}\BibitemShut {NoStop}%
\bibitem [{\citenamefont {Foerster}\ \emph {et~al.}(2017)\citenamefont
  {Foerster}, \citenamefont {Maci{\'a}}, \citenamefont {Statuto}, \citenamefont
  {Finizio}, \citenamefont {Hernandez-Minguez}, \citenamefont {Lendinez},
  \citenamefont {Santos}, \citenamefont {Fontcuberta}, \citenamefont {{Manel
  Hernandez}}, \citenamefont {Kl{\"a}ui},\ and\ \citenamefont
  {Aballe}}]{art:LNO_Ni}%
  \BibitemOpen
  \bibfield  {author} {\bibinfo {author} {\bibfnamefont {M.}~\bibnamefont
  {Foerster}}, \bibinfo {author} {\bibfnamefont {F.}~\bibnamefont {Maci{\'a}}},
  \bibinfo {author} {\bibfnamefont {N.}~\bibnamefont {Statuto}}, \bibinfo
  {author} {\bibfnamefont {S.}~\bibnamefont {Finizio}}, \bibinfo {author}
  {\bibfnamefont {A.}~\bibnamefont {Hernandez-Minguez}}, \bibinfo {author}
  {\bibfnamefont {S.}~\bibnamefont {Lendinez}}, \bibinfo {author}
  {\bibfnamefont {P.}~\bibnamefont {Santos}}, \bibinfo {author} {\bibfnamefont
  {J.}~\bibnamefont {Fontcuberta}}, \bibinfo {author} {\bibfnamefont
  {J.}~\bibnamefont {{Manel Hernandez}}}, \bibinfo {author} {\bibfnamefont
  {M.}~\bibnamefont {Kl{\"a}ui}}, \ and\ \bibinfo {author} {\bibfnamefont
  {L.}~\bibnamefont {Aballe}},\ }\href@noop {} {\bibfield  {journal} {\bibinfo
  {journal} {Nature Communications, to be published}\ } (\bibinfo {year}
  {2017})}\BibitemShut {NoStop}%
\bibitem [{\citenamefont {Liu}\ \emph {et~al.}(2013)\citenamefont {Liu},
  \citenamefont {Hoffmann}, \citenamefont {Wang}, \citenamefont {Zhang},
  \citenamefont {Nelson-Cheeseman},\ and\ \citenamefont
  {Bhattacharya}}]{art:liu_vervey_transition}%
  \BibitemOpen
  \bibfield  {author} {\bibinfo {author} {\bibfnamefont {M.}~\bibnamefont
  {Liu}}, \bibinfo {author} {\bibfnamefont {J.}~\bibnamefont {Hoffmann}},
  \bibinfo {author} {\bibfnamefont {J.}~\bibnamefont {Wang}}, \bibinfo {author}
  {\bibfnamefont {J.}~\bibnamefont {Zhang}}, \bibinfo {author} {\bibfnamefont
  {B.}~\bibnamefont {Nelson-Cheeseman}}, \ and\ \bibinfo {author}
  {\bibfnamefont {A.}~\bibnamefont {Bhattacharya}},\ }\href@noop {} {\bibfield
  {journal} {\bibinfo  {journal} {Scientific Reports}\ }\textbf {\bibinfo
  {volume} {3}},\ \bibinfo {pages} {1876} (\bibinfo {year} {2013})}\BibitemShut
  {NoStop}%
\bibitem [{\citenamefont {Adamo}\ \emph {et~al.}(2009)\citenamefont {Adamo},
  \citenamefont {Ke}, \citenamefont {Wang}, \citenamefont {Xin}, \citenamefont
  {Heeg}, \citenamefont {Hawley}, \citenamefont {Zander}, \citenamefont
  {Schubert}, \citenamefont {Schiffer}, \citenamefont {Muller}, \citenamefont
  {Maritato},\ and\ \citenamefont {Schlom}}]{art:adamo_strain_LSMO}%
  \BibitemOpen
  \bibfield  {author} {\bibinfo {author} {\bibfnamefont {C.}~\bibnamefont
  {Adamo}}, \bibinfo {author} {\bibfnamefont {X.}~\bibnamefont {Ke}}, \bibinfo
  {author} {\bibfnamefont {H.~Q.}\ \bibnamefont {Wang}}, \bibinfo {author}
  {\bibfnamefont {H.~L.}\ \bibnamefont {Xin}}, \bibinfo {author} {\bibfnamefont
  {T.}~\bibnamefont {Heeg}}, \bibinfo {author} {\bibfnamefont {M.~E.}\
  \bibnamefont {Hawley}}, \bibinfo {author} {\bibfnamefont {W.}~\bibnamefont
  {Zander}}, \bibinfo {author} {\bibfnamefont {J.}~\bibnamefont {Schubert}},
  \bibinfo {author} {\bibfnamefont {P.}~\bibnamefont {Schiffer}}, \bibinfo
  {author} {\bibfnamefont {D.~A.}\ \bibnamefont {Muller}}, \bibinfo {author}
  {\bibfnamefont {L.}~\bibnamefont {Maritato}}, \ and\ \bibinfo {author}
  {\bibfnamefont {D.~G.}\ \bibnamefont {Schlom}},\ }\href@noop {} {\bibfield
  {journal} {\bibinfo  {journal} {Applied Physics Letters}\ }\textbf {\bibinfo
  {volume} {95}},\ \bibinfo {pages} {112504} (\bibinfo {year}
  {2009})}\BibitemShut {NoStop}%
\bibitem [{\citenamefont {Roy}(2013)}]{art:roy_vortexMotionUniax}%
  \BibitemOpen
  \bibfield  {author} {\bibinfo {author} {\bibfnamefont {P.}~\bibnamefont
  {Roy}},\ }\href@noop {} {\bibfield  {journal} {\bibinfo  {journal} {Applied
  Physics Letters}\ }\textbf {\bibinfo {volume} {102}},\ \bibinfo {pages}
  {162411} (\bibinfo {year} {2013})}\BibitemShut {NoStop}%
\bibitem [{\citenamefont {Thiele}(1974)}]{art:thieleEquation}%
  \BibitemOpen
  \bibfield  {author} {\bibinfo {author} {\bibfnamefont {A.}~\bibnamefont
  {Thiele}},\ }\href@noop {} {\bibfield  {journal} {\bibinfo  {journal}
  {Physical Review Letters}\ }\textbf {\bibinfo {volume} {30}},\ \bibinfo
  {pages} {230} (\bibinfo {year} {1974})}\BibitemShut {NoStop}%
\bibitem [{\citenamefont {Huber}(1982)}]{art:huber_thieleEquation}%
  \BibitemOpen
  \bibfield  {author} {\bibinfo {author} {\bibfnamefont {D.}~\bibnamefont
  {Huber}},\ }\href@noop {} {\bibfield  {journal} {\bibinfo  {journal}
  {Physical Review B}\ }\textbf {\bibinfo {volume} {26}},\ \bibinfo {pages}
  {3758} (\bibinfo {year} {1982})}\BibitemShut {NoStop}%
\bibitem [{\citenamefont {Wysin}(1996)}]{art:wysin_thieleEquation}%
  \BibitemOpen
  \bibfield  {author} {\bibinfo {author} {\bibfnamefont {G.}~\bibnamefont
  {Wysin}},\ }\href@noop {} {\bibfield  {journal} {\bibinfo  {journal}
  {Physical Review B}\ }\textbf {\bibinfo {volume} {54}},\ \bibinfo {pages}
  {15156} (\bibinfo {year} {1996})}\BibitemShut {NoStop}%
\bibitem [{\citenamefont {{Kr\"uger}}\ \emph {et~al.}(2007)\citenamefont
  {{Kr\"uger}}, \citenamefont {Drews}, \citenamefont {Bolte}, \citenamefont
  {Merkt}, \citenamefont {Pfannkuche},\ and\ \citenamefont
  {Meier}}]{art:krueger_harmonicPotential}%
  \BibitemOpen
  \bibfield  {author} {\bibinfo {author} {\bibfnamefont {B.}~\bibnamefont
  {{Kr\"uger}}}, \bibinfo {author} {\bibfnamefont {A.}~\bibnamefont {Drews}},
  \bibinfo {author} {\bibfnamefont {M.}~\bibnamefont {Bolte}}, \bibinfo
  {author} {\bibfnamefont {U.}~\bibnamefont {Merkt}}, \bibinfo {author}
  {\bibfnamefont {D.}~\bibnamefont {Pfannkuche}}, \ and\ \bibinfo {author}
  {\bibfnamefont {G.}~\bibnamefont {Meier}},\ }\href@noop {} {\bibfield
  {journal} {\bibinfo  {journal} {Physical Review B}\ }\textbf {\bibinfo
  {volume} {76}},\ \bibinfo {pages} {224426} (\bibinfo {year}
  {2007})}\BibitemShut {NoStop}%
\bibitem [{\citenamefont {Vansteenkiste}\ \emph {et~al.}(2014)\citenamefont
  {Vansteenkiste}, \citenamefont {Leliaert}, \citenamefont {Dvornik},
  \citenamefont {Helsen}, \citenamefont {{Garcia-Sanchez}},\ and\ \citenamefont
  {{Van Waeyenberge}}}]{art:mumax}%
  \BibitemOpen
  \bibfield  {author} {\bibinfo {author} {\bibfnamefont {A.}~\bibnamefont
  {Vansteenkiste}}, \bibinfo {author} {\bibfnamefont {J.}~\bibnamefont
  {Leliaert}}, \bibinfo {author} {\bibfnamefont {M.}~\bibnamefont {Dvornik}},
  \bibinfo {author} {\bibfnamefont {M.}~\bibnamefont {Helsen}}, \bibinfo
  {author} {\bibfnamefont {F.}~\bibnamefont {{Garcia-Sanchez}}}, \ and\
  \bibinfo {author} {\bibfnamefont {B.}~\bibnamefont {{Van Waeyenberge}}},\
  }\href@noop {} {\bibfield  {journal} {\bibinfo  {journal} {AIP Advances}\
  }\textbf {\bibinfo {volume} {4}},\ \bibinfo {pages} {107133} (\bibinfo {year}
  {2014})}\BibitemShut {NoStop}%
\bibitem [{\citenamefont {Finizio}\ \emph {et~al.}(2016)\citenamefont
  {Finizio}, \citenamefont {Wintz}, \citenamefont {Kirk},\ and\ \citenamefont
  {Raabe}}]{art:bending_setup}%
  \BibitemOpen
  \bibfield  {author} {\bibinfo {author} {\bibfnamefont {S.}~\bibnamefont
  {Finizio}}, \bibinfo {author} {\bibfnamefont {S.}~\bibnamefont {Wintz}},
  \bibinfo {author} {\bibfnamefont {E.}~\bibnamefont {Kirk}}, \ and\ \bibinfo
  {author} {\bibfnamefont {J.}~\bibnamefont {Raabe}},\ }\href@noop {}
  {\bibfield  {journal} {\bibinfo  {journal} {Review of Scientific
  Instruments}\ }\textbf {\bibinfo {volume} {87}},\ \bibinfo {pages} {123703}
  (\bibinfo {year} {2016})}\BibitemShut {NoStop}%
\end{thebibliography}

%

\end{document}